\documentclass{llncs}

\usepackage{amssymb,graphicx,url,bcprules,listings,textcomp,stmaryrd,xypic}

\newcommand{\qq}{\texttt{"}}

\lstdefinelanguage{LambdaJS}{
sensitive=false,
morekeywords={let,func,return,undefined,null,typeof,true,false,ref,deref,
              var,new,break,while,try,catch,finally,throw,err,if,else,for,do,
              with,delete,new,instanceof,},
keywordstyle=\bfseries\sffamily,
identifierstyle=\sffamily,
comment=[l]{//},
commentstyle=\sffamily,
string=[d]{"},
stringstyle=\sffamily,
mathescape=true,
extendedchars=true,
basicstyle=\footnotesize\ttfamily,
showstringspaces=false,
numbers=none,
firstnumber=0,
numberstyle=\tiny,
stepnumber=5,
numbersep=5pt,
upquote=true,
columns=fullflexible,
flexiblecolumns=true,
}

\lstdefinelanguage{JavaScript}{
sensitive=false,
morekeywords={let,function,return,undefined,null,typeof,true,false,ref,deref,
              var,new,break,while,try,catch,finally,throw,err,if,else,for,do,
              with,delete,new,instanceof,this,},
keywordstyle=\ttfamily,
identifierstyle=\ttfamily,
comment=[l]{//},
commentstyle=\ttfamily,
string=[d]{"},
stringstyle=\ttfamily,
mathescape=true,
extendedchars=true,
basicstyle=\footnotesize\ttfamily,
showstringspaces=false,
numbers=none,
firstnumber=0,
numberstyle=\tiny,
stepnumber=5,
numbersep=5pt,
upquote=true,
columns=fixed,
flexiblecolumns=true,
}
\def\js#1{\textrm{\lstinline|#1|}}

\def\refsect#1{section~\ref{#1}}

\def\reffig#1{figure~\ref{#1}}
\def\Reffig#1{Figure~\ref{#1}}
\def\refthm#1{theorem~\ref{#1}}

\def\refdef#1{definition~\ref{#1}}

\def\reflem#1{lemma~\ref{#1}}

\def\refprop#1{proposition~\ref{#1}}

\def\XMLHttpRequest{{\sc XMLHttpRequest}}
\def\FBJS{{\sc fbjs}}

\def\JS{\textbf{JS}}
\def\Safe{\textbf{NotXHR}}
\def\hole{\bullet}

\newtheorem{safety1}{Definition}
\newtheorem{desugarclaim1}{Claim}
\newtheorem{desugarclaim2}[desugarclaim1]{Claim}
\newtheorem{safety1thm}{Theorem}
\newtheorem{progress}{Property}
\newtheorem{safety1lem1}{Lemma}
\newtheorem{safety1lem2}[safety1lem1]{Lemma}
\newtheorem{replacement}[safety1lem1]{Lemma}
\newtheorem{plusreplacement}[safety1lem1]{Lemma}

\newtheorem{prop1}{Proposition}

\begin{document}

\mainmatter

\title{The Essence of JavaScript}

\author{
Arjun Guha \and
Claudiu Saftoiu \and
Shriram Krishnamurthi}
\institute{Brown University}

\maketitle

\newlength{\bannerlen}
\settowidth{\bannerlen}{\mbox{Corrected: October 3, 2015}}
\newlength{\bannerspc}
\setlength{\bannerspc}{0.5\textwidth}
\addtolength{\bannerspc}{-0.5\bannerlen}

\vbox to 0pt{%
\vskip -120pt
\hspace{\bannerspc}\textbf{Corrected: October 3, 2015}\hspace{\bannerspc}
\vss}%

\begin{abstract}
We reduce JavaScript to a core calculus structured as a small-step
operational semantics.  We present several peculiarities of the
language and show that our calculus models them.  We explicate the
desugaring process that turns JavaScript programs into ones in the
core.  We demonstrate faithfulness to JavaScript using real-world test
suites.  Finally, we illustrate utility by defining a security
property, implementing it as a type system on the core, and extending
it to the full language.
\end{abstract}

\section{The Need for Another JavaScript Semantics}

The growing use of JavaScript
has created whole new technical and business models
of program construction and deployment.
JavaScript is a feature-rich language with many quirks,
and these quirks are often exploited by security and privacy
attacks.  This is especially true in cases where JavaScript has a
familiar syntax but an unconventional semantics.

Due to its popularity and shortcomings, companies and researchers have
tried to tame JavaScript via program
analyses~\cite{chugh:sif,livshits:gatekeeper,guha:intrusiondetection,jensen:tajs},
sub-language~\cite{adsafe,fbjs,miller:caja}, and more.  These
works claim but do not demonstrate soundness, partly because we lack
a tractable account of the language.
The JavaScript
standard~\cite{ecma262} is capacious and informal, while one major
formal semantics~\cite{maffeis:jssemantics} is large, not amenable to
conventional proof techniques, and inherits the standard's
complexities, as we discuss in
\refsect{s:related}.  In contrast:
\begin{itemize}

\item
We present a core language, $\lambda_{JS}$,
that embodies JavaScript's essential features
 (sans \lstinline|eval|).
$\lambda_{JS}$ fits on three pages and lends itself well to
 proof techniques such as subject reduction.

\item
 We show that we can desugar JavaScript into $\lambda_{JS}$.
 In particular, desugaring handles notorious JavaScript features
 such as \lstinline|this| and
 \lstinline|with|, so $\lambda_{JS}$
 itself remains simple (and thus simplifies proofs that utilize it).

\item
 We mechanize both $\lambda_{JS}$ and desugaring.

\item
 To show compliance with reality, we successfully test
  $\lambda_{JS}$ and desugaring against the actual
  Mozilla
  JavaScript test suite.

\item
 Finally, we demonstrate the use of our semantics by building a safe
 subset of JavaScript.  This application highlights how our
 partitioning of JavaScript into core and syntactic sugar lends
 structure to proofs.
\end{itemize}
Our supplemental materials (full desugaring, tools, etc.) are
available at\newline
\begin{center}
\url{http://www.cs.brown.edu/research/plt/dl/jssem/v1/}
\end{center}

\section{$\lambda_{JS}$: A Tractable Semantics for JavaScript}
\label{s:lambdaJS}

JavaScript is full of surprises.  Syntax that may have a conventional
interpretation for many readers often has a subtly different semantics
in JavaScript. To aid the reader, we introduce $\lambda_{JS}$
incrementally. We include examples of JavaScript's quirks and show how
$\lambda_{JS}$ faithfully models them.

Figures~\ref{f:functional}, \ref{f:state}, \ref{f:prototypes},
\ref{f:control}, and \ref{f:prims} specify the syntax and semantics of
$\lambda_{JS}$.  We use a Felleisen-Hieb small-step operational semantics with
evaluation contexts~\cite{felleisen:redex}. We typeset $\lambda_{JS}$ code in a
\textsf{sans-serif typeface}, and JavaScript in a \texttt{fixed-width typeface}.

\subsection{Functions, Objects and State}
\label{s:functional}
\lstset{language=LambdaJS}

\begin{figure}[t]
\begin{displaymath}
\begin{array}{rcl}
c & = & 
\js{$num \mid str \mid bool \mid\ $ undefined $\ \mid\ $ null} \\
v &= &  c \mid \js{func($x\cdots$) \{ return\ $e\ $ \}} \ \mid\ 
         \js{\{\ $str$ : $v$ $\cdots\ $ \}} \\
e & = & x \mid v \mid
        \js{let ($x\ $ =\ $e$)\ $e$ $\ \mid\ $ $e$($e\cdots$)} 
 \mid \js{$e$[$e$]}
 \mid \js{$e$[$e$] =\ $e$}
 \mid \js{delete\ $e$[$e$]} \\
E & =    & \hole
    \mid   \js{let ($x\ $ =\ $E$)\ $e$}
    \mid   \js{$E$($e\cdots$)}
    \mid   \js{$v$($v\cdots\  E$,\ $e\cdots$)} \\
  & \mid & \js{ \{ $str$:$\ v \cdots\ str$: $E$,\ $str$: $e\cdots\ $\} }
   \mid  \js{$E$[$e$]} \mid \js{$v$[$E$]} \mid 
   \js{$E$[$e$] =\ $e$}  \mid   \js{$v$[$E$] =\ $e$} \\
  & \mid & \js{$v$[$v$] =\ $E$}
    \mid   \js{delete\ $E$[$e$]}
    \mid   \js{delete\ $v$[$E$]}
  
\end{array}
\end{displaymath}

\infax[E-Let]
{\js{let ($x\ $ =\ $v$)\ $e$} \hookrightarrow e[x/v]}

\infax[E-App]
{\js{(func($x_1\cdots x_n$) \{ return\ $e\ $ \})($v_1\cdots v_n$)}
 \hookrightarrow e[x_1/v_1 \cdots x_n/v_n]}

\infax[E-GetField]
{\js{\{ $\ \cdots str$:\ $v \cdots\ $ \}[$str$]}
 \hookrightarrow v}

\infrule[E-GetField-NotFound]
{str_x \not\in  (str_1 \cdots str_n)}
{\js{\{$\ str_1$ : $\ v_1\  \cdots\ str_n$ : $\ v_n\ $\}
 [$str_x$]}
 \hookrightarrow \js{undefined}}

\infax[E-UpdateField]
{\js{\{$\ str_1$: $\ v_1\cdots\ str_i$ : 
       $\ v_i \ \cdots str_n$ : $\ v_n\ $\} [$str_i$] =\ $v$} \\
 \hookrightarrow
{\js{\{$\ str_1$: $\ v_1\cdots\ str_i$ : 
       $\ v \ \cdots str_n$ : $\ v_n\ $\}}}}

\infrule[E-CreateField]
{str_x \not\in (str_1 \cdots)}
{\js{\{$\ str_1$: $\ v_1\cdots\ $ \} [$str_x$] = $\ v_x$}
 \hookrightarrow
 \js{\{$\ str_x$: $\ v_x$, $\ str_1$: $\ v_1\cdots\ $ \}}}

\infax[E-DeleteField]
{\js{delete \{$\ str_1$: $\ v_1\cdots\ str_x$ : 
       $\ v_x \ \cdots str_n$ : $\ v_n\ $\} [$str_x$]} \\
 \hookrightarrow
{\js{\{$\ str_1$: $\ v_1\cdots str_n$ : $\ v_n\ $\}}}}

\infrule[E-DeleteField-NotFound]
{str_x \not\in (str_1\cdots)}
{\js{delete \{$\ str_1$: $\ v_1\cdots\ $\} [$str_x$]}
 \hookrightarrow
 \js{\{$\ str_1$: $\ v_1\cdots\ $\}}}

\medskip\hrule
\caption{Functions and Objects}
\label{f:functional}
\end{figure}

\begin{figure}[t]

\begin{displaymath}
\begin{array}{rclr}
l & = & \cdots & \textrm{Locations} \\
v & = & \cdots \mid l & \textrm{Values} \\
\sigma & = & (l,v)\cdots & \textrm{Stores} \\
e & =    & \cdots 
    \mid   \js{$e\ $ =\ $e$}
    \mid   \js{ref\ $e$}
    \mid   \js{deref\ $e$}
& \textrm{Expressions} \\
E & =   & \cdots
   \mid   \js{$E\ $ =\ $e$}
   \mid   \js{$v\ $ =\ $E$}
   \mid   \js{ref\ $E$}
   \mid   \js{deref\ $E$}
& \textrm{\ \ \ \ \ Evaluation Contexts}
\end{array}
\end{displaymath}

\infrule
{e_1 \hookrightarrow e_2}
{\sigma E\langle e_1 \rangle \rightarrow \sigma E\langle e_2 \rangle}

\infrule[E-Ref]
{l \not\in dom(\sigma) \andalso \sigma' = \sigma,(l, v)}
{\sigma E\langle \js{ref\ $v$}\rangle \rightarrow
 \sigma' E\langle l \rangle}

\infax[E-Deref]
{\sigma E\langle \js{deref\ $l$}\rangle \rightarrow
 \sigma E\langle \sigma(l) \rangle}

\infax[E-SetRef]
{\sigma E\langle \js{$l\ $ =\ $v$}\rangle \rightarrow
 \sigma\lbrack l := v\rbrack E\langle v \rangle,\textrm{ if $l \in dom(\sigma)$}}

We use $\twoheadrightarrow$ to denote the reflexive-transitive closure
of $\rightarrow$.

\medskip\hrule
\caption{Mutable References in $\lambda_{JS}$}
\label{f:state}
\end{figure}

\lstset{language=JavaScript}

\begin{figure}[t]
\begin{lstlisting}
function sum(arr) {
  var r = 0;
  for (var i = 0; i < arr["length"]; i = i + 1) {
    r = r + arr[i] };
  return r };

sum([1,2,3]) $\twoheadrightarrow$ 6
var a = [1,2,3,4];
delete a["3"];
sum(a) $\twoheadrightarrow$ NaN
\end{lstlisting}
\medskip\hrule
\caption{Array Processing in JavaScript}
\label{f:arrays}
\end{figure}

\lstset{language=LambdaJS}

We begin with the small subset of $\lambda_{JS}$ specified in
\reffig{f:functional} that includes just functions and objects.  We
model operations on objects via functional update.  This seemingly
trivial fragment already exhibits some of JavaScript's quirks:
\begin{itemize}

\item
In field lookup, the name of the field need not be specified
statically; instead, field names may be computed at runtime
(\textsc{E-GetField}):
\begin{lstlisting}
let (obj = { $\qq$x$\qq$ : 500, $\qq$y$\qq$ : 100 })
  let (select = func(name) { return obj[name] })
    select($\qq$x$\qq$) + select($\qq$y$\qq$) 
$\hookrightarrow^*$ 600
\end{lstlisting}

\item
A program that looks up  a non-existent field does not
  result in an error; instead, JavaScript returns the value
  \lstinline|undefined| (\textsc{E-GetField-NotFound}):
\begin{lstlisting}
{ $\qq$x$\qq$ : 7 }[$\qq$y$\qq$] $\hookrightarrow$ undefined
\end{lstlisting}

\item Field update in JavaScript is conventional (\textsc{E-UpdateField})---
\begin{lstlisting}
{ $\qq$x$\qq$ : 0 }[$\qq$x$\qq$] = 10 $\hookrightarrow$ { $\qq$x$\qq$ : 10 }
\end{lstlisting}
---but the same syntax also creates new fields
(\textsc{E-CreateField}):
\begin{lstlisting}
{ $\qq$x$\qq$ : 0 }[$\qq$z$\qq$] = 20 $\hookrightarrow$ {$\qq$z$\qq$ : 20, $\qq$x$\qq$ : 10 }
\end{lstlisting}

\item Finally, JavaScript lets us delete fields from objects:
\begin{lstlisting}
delete { $\qq$x$\qq$: 7, $\qq$y$\qq$: 13}[$\qq$x$\qq$] $\hookrightarrow$ { $\qq$y$\qq$: 13 }
\end{lstlisting}
\end{itemize}

\lstset{language=JavaScript}
JavaScript also supports a more conventional dotted-field
notation: \lstinline|obj.x| is valid JavaScript, and is
equivalent to \lstinline|obj[$\qq$x$\qq$]|.  To keep
$\lambda_{JS}$ small, we omit the dotted-field notation in favor of
the more general computed lookup, and instead explicitly treat dotted
fields as syntactic sugar.

\subsubsection{Assignment and Imperative Objects}
\lstset{language=JavaScript}
JavaScript has two forms of state: objects are mutable, and variables
are assignable.  We model both variables and imperative objects with
first-class mutable 
references (\reffig{f:state}).\footnote{In the semantics, we use 
$E\langle e \rangle$ instead of the conventional
$E[e]$ to denote a filled evaluation context, to avoid confusion with 
JavaScript's objects.} We desugar
JavaScript to explicitly allocate and dereference heap-allocated values
in $\lambda_{JS}$.

\paragraph{Example: JavaScript Arrays}
JavaScript has arrays that developers tend to use in a
traditional imperative style.  However, JavaScript arrays are really
objects, and this can lead to unexpected behavior.
\Reffig{f:arrays} shows a small example of a seemingly conventional
 use of arrays.
Deleting the field \lstinline|a["3"]| (\textsc{E-DeleteField})
does not affect \lstinline|a["length"]| or shift the
array elements.  Therefore, in the loop body, \lstinline|arr["3"]|
evaluates to \lstinline|undefined|, via \textsc{E-GetField-NotFound}.
Finally, adding \lstinline|undefined| to a number yields \lstinline|NaN|;
we discuss other quirks of addition in \refsect{s:prims}.

\lstset{language=LambdaJS}

\begin{figure}
\infrule[E-GetField-NotFound]
{str_x \notin  (str_1 \cdots str_n) \andalso
 \js{$\qq$__proto__$\qq$} \not\in (str_1\cdots str_n)}
{\js{\{$\ str_1\ $ : $\ v_1\ $, $\ \cdots\ $, $\ str_n\ $ : $\ v_n\ $\}
 [$str_x$]}
 \hookrightarrow \js{undefined}}

\infrule[E-GetField-Proto-Null]
{str_x \notin  (str_1 \cdots str_n)}
{\js{\{$\ str_1\ $ : $\ v_1 \cdots\ $ $\qq$__proto__$\qq$: null $\ \cdots$
  $\ str_n\ $ : $\ v_n\ $\}
 [$str_x$]}
 \hookrightarrow \js{undefined}}

\infrule[E-GetField-Proto]
{str_x \notin  (str_1 \cdots str_n)}
{\js{\{$\ str_1\ $ : $\ v_1 \cdots\ $ $\qq$__proto__$\qq$: $\ l$ $\ \cdots$
  $\ str_n\ $ : $\ v_n\ $\}
 [$str_x$]}
 \hookrightarrow \js{(deref $\ l$)[$str_x$]}}

\medskip\hrule
\caption{Prototype-Based Objects}
\label{f:prototypes}
\end{figure}

\subsection{Prototype-Based Objects}
\label{s:prototypes}

\lstset{language=JavaScript}
JavaScript supports \emph{prototype
  inheritance}~\cite{borning:prototypes}.  For example, in the following
code, \js{animal} is the prototype of \js{dog}:
\begin{lstlisting}
var animal = { "length": 13, "width": 7 };
var dog = { "__proto__": animal, "barks": true };
\end{lstlisting}
Prototypes affect field lookup:
\begin{lstlisting}
dog["length"] $\twoheadrightarrow$ 13
dog["width"]  $\twoheadrightarrow$ 7

var lab = { "__proto__": dog, "length": 2 }
lab["length"] $\twoheadrightarrow$ 2
lab["width"] $\twoheadrightarrow$ 7
lab["barks"] $\twoheadrightarrow$ true
\end{lstlisting}

Prototype inheritance does not affect field update.  The code below
creates the field \js{dog["width"]}, but it does not affect
\js{animal["width"]}, which \js{dog} had previously inherited:
\begin{lstlisting}
dog["width"] = 19
dog["width"] $\twoheadrightarrow$ 19
animal["width"] $\twoheadrightarrow$ 7
\end{lstlisting}
However, \js{lab} now inherits \js{dog["width"]}:
\begin{lstlisting}
lab["width"] $\twoheadrightarrow$ 19
\end{lstlisting}

\lstset{language=LambdaJS}
\Reffig{f:prototypes} specifies prototype inheritance.
The figure  modifies \textsc{E-GetField-NotFound} to only apply when the
\lstinline|$\qq$__proto__$\qq$| field is missing.

Prototype inheritance is simple, but it is obfuscated by JavaScript's
syntax.  The examples in this section are not standard
JavaScript because the \lstinline|$\qq$__proto__$\qq$| field is not directly
accessible by JavaScript programs.\footnote{Some browsers, such as
Firefox, can run these examples.}
  In the next section, we unravel and
desugar JavaScript's syntax for prototypes.

\subsection{Prototypes}
\label{s:proto-stx}

\begin{figure}
\lstset{language=JavaScript}
$desugar\llbracket$\lstinline|{$prop$: $e\cdots$}|
$\rrbracket =$
\lstset{language=LambdaJS}\begin{lstlisting}
ref { 
  prop : $desugar\llbracket e\rrbracket\cdots$, 
  $\qq$__proto__$\qq$: (deref Object)[$\qq$prototype$\qq$] 
}
\end{lstlisting}

\lstset{language=JavaScript}
$desugar\llbracket$\lstinline|function($x\cdots$) { $stmt\cdots$ }|
$\rrbracket =$
\lstset{language=LambdaJS}\begin{lstlisting}
 ref { 
  $\qq$code$\qq$: func(this, $x\cdots$) { return $desugar\llbracket stmt\cdots \rrbracket$ },
  $\qq$prototype$\qq$: ref { $\qq$__proto___$\qq$: (deref Object)[$\qq$prototype$\qq$] } }
\end{lstlisting}

\lstset{language=JavaScript}
$desugar\llbracket$\lstinline|new $e_f$($e\cdots$)}|$\rrbracket =$
\lstset{language=LambdaJS}
\begin{lstlisting}
  let (constr = deref $desugar\llbracket e_f \rrbracket$)
    let (obj = ref { $\qq$__proto__$\qq$ : constr[$\qq$prototype$\qq$]})
      constr[$\qq$code$\qq$](obj, $desugar\llbracket e\rrbracket \cdots$);
      obj
\end{lstlisting}

\lstset{language=JavaScript}
$desugar\llbracket$\lstinline|$obj$[$field$]($e\cdots$)|$\rrbracket = $
\lstset{language=LambdaJS}
\begin{lstlisting}
  let (obj = $desugar\llbracket obj \rrbracket$)
    let (f = (deref obj)[field])
      f[$\qq$code$\qq$](obj, $desugar\llbracket e\rrbracket\cdots$)
\end{lstlisting}

\lstset{language=JavaScript}
$desugar\llbracket$\lstinline|$e_f$($e\cdots$)|$\rrbracket = $
\lstset{language=LambdaJS}
\begin{lstlisting}
  let (obj = $desugar\llbracket e_f \rrbracket$)
    let (f = deref obj)
      f[$\qq$code$\qq$](window, $desugar\llbracket e\rrbracket\cdots$)
\end{lstlisting}

\lstset{language=JavaScript}
$desugar\llbracket$\lstinline|$obj$ instanceof $constr$|$\rrbracket =$
\lstset{language=LambdaJS}
\begin{lstlisting}
  let (obj = ref (deref $desugar\llbracket obj \rrbracket$),
       constr = deref $desugar\llbracket constr \rrbracket$)
    done: {
      while (deref obj !== null) {
        if ((deref obj)[$\qq$__proto__$\qq$] === constr[$\qq$prototype$\qq$]) {
          break done true }
        else { obj = (deref obj)[$\qq$__proto__$\qq$] } };
      false }
\end{lstlisting}

\lstset{language=JavaScript}
$desugar\llbracket$\lstinline|this|$\rrbracket = $
\lstset{language=LambdaJS}
\lstinline|this|
(an ordinary identifier, bound by functions)

\lstset{language=JavaScript}
$desugar\llbracket$\lstinline|$e$.$x$|$\rrbracket = $
\lstset{language=LambdaJS}
\lstinline|$desugar\llbracket e\rrbracket$[$\qq x\qq$]|

\medskip\hrule
\caption{Desugaring JavaScript's Object Syntax}
\label{f:classes}
\end{figure}

\lstset{language=JavaScript}
JavaScript programmers can indirectly manipulate prototypes
using syntax that is reminiscent of class-based languages like
Java.  In this section, we explain this syntax and its
actual semantics.  
We account for this class-like
syntax by desugaring it to manipulate prototypes directly
(\refsect{s:prototypes}).  Therefore, this section does not grow
$\lambda_{JS}$ and only describes desugaring.
  \Reffig{f:classes} specifies the portion of desugaring
that is relevant for the rest of this section.

\subsubsection{The \texttt{this} Keyword}

\begin{figure}[t]
\begin{lstlisting}
var obj = {
  "x" : 0,
  "setX": function(val) { this.x = val } };

// window is the name of the global object in Web browsers
window.x $\twoheadrightarrow$ undefined
obj.setX(10);
obj.x  $\twoheadrightarrow$ 10
var f = obj.setX;
f(90);
obj.x  $\twoheadrightarrow$ 10 // obj.x was not updated
window.x $\twoheadrightarrow$ 90 // window.x was created
\end{lstlisting}
\medskip\hrule
\caption{Implicit \lstinline|this| Parameter}
\label{f:implicitthis}
\end{figure}

JavaScript does not have conventional methods.  Function-valued fields
are informally called ``methods'', and provide an interpretation for a
\lstinline|this| keyword, but both are quite different from those of,
say, Java.

For example, in \reffig{f:implicitthis}, when \lstinline|obj.setX(10)|
is applied, \lstinline|this| is bound to \lstinline|obj| in the body of
the function. In the same figure however, although \lstinline|f| is
bound to \lstinline|obj.setX|, \lstinline|f(90)| does not behave like a
traditional method call.  In fact, the function is applied with
\lstinline|this| bound to the \emph{global
  object}~\cite[Section 10.1.5]{ecma262}.

In general, \lstinline|this| is an implicit parameter to all JavaScript
functions.  Its value is determined by the syntactic shape of function
applications.  Thus, when we desugar functions to $\lambda_{JS}$, we make
\lstinline|this| an explicit argument. Moreover, we desugar function
calls to explicitly supply a value for \lstinline|this|.

\subsubsection{Functions as Objects}

In JavaScript, functions are objects with fields:
\begin{lstlisting}
f = function(x) { return x + 1 }
f.y = 90
f(f.y) $\twoheadrightarrow$ 91
\end{lstlisting}
We desugar JavaScript's \lstinline|function| to objects in
$\lambda_{JS}$ with a distinguished \lstinline|code| field that
refers to the actual function.
  Therefore, we also desugar application to lookup
the \lstinline|code| field.

We could design $\lambda_{JS}$ so that functions truly are objects,
making this bit of desugaring unnecessary.  In our experience,
JavaScript functions are rarely used as objects.  Therefore, our design
lets us reason about simple functions when possible, and functions as
objects only when necessary.

In addition to the \lstinline|code| field, which we add by desugaring,
and any other fields that may have been created by the
programmer, all functions also have a distinguished field called
\lstinline|prototype|.  As
\reffig{f:classes} shows, the \lstinline|prototype| field is a reference 
to an object that
eventually leads to the prototype of \lstinline|Object|.  Unlike the
\lstinline|__proto__| field, \lstinline|prototype| is accessible and
can be updated by programmers. The combination of its mutability and
its use in \lstinline|instanceof| leads to unpredictable behavior,
as we show below.

\subsubsection{Constructors and Prototypes}

JavaScript does not have explicit constructors, but it does have a
\lstinline|new| keyword that invokes a function with
\lstinline|this| bound to a new object.  For example, the following
code---
\begin{lstlisting}
function Point(x, y) {
  this.x = x;
  this.y = y }

pt = new Point(50, 100)
\end{lstlisting}
---applies the function \lstinline|Point| and returns the value of 
\lstinline|this|.
\lstinline|Point| explicitly sets \lstinline|this.x| and \lstinline|this.y|.
 Moreover, \lstinline|new Point| implicitly sets \lstinline|this.__proto__| to
\lstinline|Point.prototype|.
We can now observe prototype inheritance:
\begin{lstlisting}
Point.prototype.getX = function() { return this.x }
pt.getX() $\twoheadrightarrow$ pt.__proto__.getX() $\twoheadrightarrow$ 50
\end{lstlisting}
In standard JavaScript, because the \lstinline|__proto__| field is not
exposed, the only way to set up a prototype hierarchy is to update
the \lstinline|prototype| field of functions that are used as constructors.

\subsubsection{The \lstinline|instanceof| Operator}

\begin{figure}[t]
\begin{lstlisting}
function Dog() { this.barks = "woof" };
function Cat() { this.purrs = "meow" };
dog = new Dog();
cat = new Cat();
dog.barks; $\twoheadrightarrow$ "woof"
cat.purrs; $\twoheadrightarrow$ "meow"

function animalThing(obj) {
  if (obj instanceof Cat) { return obj.purrs }
  else if (obj instanceof Dog) { return obj.barks }
  else { return "unknown animal" } };

animalThing(dog); $\twoheadrightarrow$ "woof"
animalThing(cat); $\twoheadrightarrow$ "meow" 
animalThing(4234); $\twoheadrightarrow$ "unknown animal"

Cat.prototype = Dog.prototype;
animalThing(cat); $\twoheadrightarrow$ "unknown animal"
animalThing(dog) $\twoheadrightarrow$ undefined // dog.purrs ($\textrm{\textsc{E-GetField-NotFound}}$)
\end{lstlisting}
\medskip\hrule
\caption{Using \lstinline|instanceof|}
\label{f:instanceof}
\end{figure}

JavaScript's \lstinline|instanceof| operator has an unconventional
semantics that reflects the peculiar notion of constructors
that we have already discussed.  In most languages, a programmer might
expect that if \lstinline|x| is bound to the value created by 
\lstinline|new Constr($\cdots$)|, then
\lstinline|x instanceof Constr| is true.  In JavaScript, however, 
this invariant does not apply.

For example, in \reffig{f:instanceof},
\lstinline|animalThing| dispatches on the type of its argument using
\lstinline|instanceof|.  However, after we set 
\lstinline|Cat.prototype = Dog.prototype|,
 the type structure seems to break down. The resulting behavior might
 appear unintuitive in JavaScript, but it is straightforward when we
 desugar \lstinline|instanceof| into $\lambda_{JS}$.  In essence,
\lstinline|cat instanceof Cat| is 
\lstinline|cat.__proto__ === Cat.prototype|.\footnote{The
\lstinline|===| operator is the physical equality operator, akin to
\lstinline|eq?| in Scheme.}
In the figure, before \lstinline|Cat.prototype = Dog.prototype| is
evaluated, the following are true:
\begin{lstlisting}
cat.__proto__ === Cat.prototype
dog.__proto__ === Dog.prototype
Cat.prototype !== Dog.prototype
\end{lstlisting}
However, after we update \lstinline|Cat.prototype|, we have:
\begin{lstlisting}
cat.__proto__ === $\textrm{the previous value of}$ Cat.prototype
dog.__proto__ === Dog.prototype
Cat.prototype === Dog.prototype
\end{lstlisting}
Hence, \lstinline|cat instanceof Cat| becomes \lstinline|false|.
Furthermore, since \lstinline|animalThing| first tests for
\lstinline|Cat|, the test \lstinline|dog instanceof Cat| succeeds.

\subsection{Statements and Control Operators}
\label{s:control}
\lstset{language=LambdaJS}

\begin{figure}
\begin{displaymath}
\begin{array}{rcl}
label & = & \textrm{(Labels)} \\
e & =    & \cdots 
    \mid   \js{if ($e$) \{\ $e\ $ \} else \{\ $e\ $ \}}
    \mid   \js{$e$;$e$}
    \mid   \js{while($e$) \{\ $e\ $ \}}
    \mid   \js{$label$:\{\ $e\ $ \}} \\
  & \mid & \js{break\ $label\ $ $e$}
    \mid   \js{try \{\ $e\ $ \} catch ($x$) \{\ $e\ $ \}}
    \mid   \js{try \{\ $e\ $ \} finally \{\ $e\ $ \}} \\
  & \mid & \js{err\ $v$} 
    \mid   \js{throw\ $e$} \\
E & =   & \cdots
    \mid  \js{if ($E$) \{\ $e\ $ \} else \{\ $e\ $ \}}
    \mid  \js{$E$;$e$}
    \mid  \js{$label$:\{\ $E\ $ \}}
    \mid  \js{break\ $label\ $ $E$} \\
  & \mid & \js{try \{\ $E\ $ \} catch ($x$) \{\ $e\ $ \}}
    \mid   \js{try \{\ $E\ $ \} finally \{\ $e\ $ \}}
    \mid   \js{throw\ $E$} \\
E' & =    & \hole \mid
  \js{let ($x\ $ =\ $v\cdots\ $ $x\ $ =\ $E^\prime$,\ $x\ $ =\ $e\cdots$)\ $e$}
    \mid   \js{$E^\prime$($e\cdots$)}
    \mid   \js{$v$($v\cdots\ E^\prime$,\ $e\cdots$)} \\
  & \mid &  \js{if ($E^\prime$) \{\ $e\ $ \} else \{\ $e\ $ \} }
    \mid   \js{ \{\ $str$:\ $v\cdots\ str$:\ $E^\prime$,\ $str$:\ $e\cdots\ $\} } \\
  & \mid & \js{$E^\prime$[$e$]}
    \mid   \js{$v$[$E^\prime$]}
    \mid   \js{$E^\prime$[$e$] =\ $e$}
    \mid   \js{$v$[$E^\prime$] =\ $e$}
    \mid   \js{$v$[$v$] =\ $E^\prime$}
    \mid   \js{$E^\prime\ $ =\ $e$}
    \mid   \js{$v\ $ =\ $E^\prime$} \\
  & \mid & \js{delete\ $E^\prime$[$e$]}
    \mid   \js{delete\ $v$[$E^\prime$]}
    \mid   \js{ref\ $E^\prime$}
    \mid   \js{deref\ $E^\prime$}
    \mid   \js{$E^\prime$;\ $e$} 
    \mid   \js{throw\ $E^\prime$} \\
F & = & E' \mid \js{$label$:\{\ $F\ $ \}}
    \mid  \js{break\ $label\ $ $F$} \textrm{\ \ \ (Exception Contexts)} \\
G & = & E' \mid \js{try \{\ $G\ $ \} catch ($x$) \{\ $e\ $ \}} 
\textrm{\ \ \ (Local Jump Contexts)}
\end{array}
\end{displaymath}

\infax[E-IfTrue]
{\js{if (true) \{\ $e_1\ $\} else \{\ $e_2\ $\}} \hookrightarrow e_1}

\infax[E-IfFalse]
{\js{if (false) \{\ $e_1\ $\} else \{\ $e_2\ $\}} \hookrightarrow e_2}

\infax[E-Begin-Discard]
{\js{$v$;$e$ $\ \hookrightarrow\ $ $e$}}

\infax[E-While]
{\js{while($e_1$) \{\ $e_2\ $ \}}
 \hookrightarrow
  \js{if ($e_1$) \{\ $e_2$; while($e_1$) \{\ $e_2\ $ \} \} else \{ undefined \}}}

\infax[E-Throw]
{\js{throw $\ v$}
  \hookrightarrow
  \js{err $\ v$}}

\infax[E-Catch]
{\js{try \{\ $F\langle$err $\ v$$\rangle\ $ \} catch ($x$) \{\ $e\ $ \}}
 \hookrightarrow
  e[x/v]}

\infax[E-Uncaught-Exception]
{\sigma F\langle\js{err\ $v$}\rangle \rightarrow \sigma \js{err\ $v$}}

\infax[E-Finally-Error]
{\js{try \{\ $F\langle$err $\ v$$\rangle\ $ \} finally \{\ $e\ $ \}}
 \hookrightarrow
 \js{$e$; err $\ v$}}

\infax[E-Finally-Break]
{\js{try \{\ $G\langle$break $\ label\ v$$\rangle\ $ \} finally \{\ $e\ $ \}}
 \hookrightarrow
 \js{$e$; break $\ label\ v$}}

\infax[E-Catch-Pop]
{\js{try \{ $\ v\ $ \} catch ($x$) \{ $\ e\ $ \}}
 \hookrightarrow
 \js{$v$}}

\infax[E-Finally-Pop]
{\js{try \{ $\ v\ $ \} finally \{ $\ e\ $ \}}
 \hookrightarrow
 \js{$e$; $\ v$}}

\infax[E-Break]
{\js{$label$:\{\ $G\langle$break $\ label$ $\ v$$\rangle\ $ \}}
 \hookrightarrow
 v}

\infrule[E-Break-Pop]
{label_1 \ne label_2}
{\js{$label_1$:\{\ $G\langle$break $\ label_2\ v$$\rangle\ $ \}}
 \hookrightarrow
 \js{break $\ label_2\ v$}}

\infax[E-Label-Pop]{\js{$label$:\{ $\ v\ $ \}} \hookrightarrow v}

\infax[E-Break-Break]
{\js{break\ $label_1\ G\langle$break\ $label_2\ $\  $v\rangle$} 
 \hookrightarrow
 \js{break\ $label_2\ $$v$}}

\medskip\hrule
\caption{Control operators for $\lambda_{JS}$}
\label{f:control}
\end{figure}

JavaScript has a plethora of control statements.  Many map directly to
$\lambda_{JS}$'s control operators (\reffig{f:control}), while the rest
are easily desugared.

\lstset{language=JavaScript}
For example, consider JavaScript's \lstinline|return| and
\lstinline|break| statements.  A \lstinline|break $l$| statement
transfers control to the local label $l$.  A \lstinline|return $e$|
statement transfers control to the end of the local function and
produces the value of $e$ as the result.
\lstset{language=LambdaJS}
Instead of two control operators
that are almost identical, $\lambda_{JS}$ has a single \lstinline|break|
expression that produces a value.

Concretely, we elaborate JavaScript's functions to begin with a label
\lstinline|ret|:
\begin{center}
\lstset{language=JavaScript}
$desugar\llbracket$\lstinline|function($x\cdots$) { $stmt\cdots$ }|
$\rrbracket = $\lstset{language=LambdaJS}
\lstinline|func($this\ x\cdots$) { return $ret$: { $desugar\llbracket stmt\cdots\rrbracket$ } }|
\end{center}
Thus, \texttt{return} statements are desugared to \lstinline|break ret|:
\begin{center}
\lstset{language=JavaScript}
$desugar\llbracket$\lstinline|return $e$|$\rrbracket = $
\lstset{language=LambdaJS}\lstinline|break ret $desugar\llbracket e\rrbracket$|
\end{center}
while \texttt{break} statements are desugared to produce
\lstinline|undefined|:
\begin{center}
\lstset{language=JavaScript}
$desugar\llbracket$\lstinline|break $label$|$\rrbracket = $
\lstset{language=LambdaJS}\lstinline|break $label$ undefined|
\end{center}

\subsection{Static Scope in JavaScript}
\label{s:scope}

\lstset{language=JavaScript}
The JavaScript standard specifies identifier lookup in an unconventional
manner. It uses neither substitution nor environments, but 
\emph{scope objects}~\cite[Section 10.1.4]{ecma262}.
A scope object is akin to an
activation record, but is a conventional JavaScript object.  The fields of
this object are interpreted as variable bindings.

In addition, a scope object has a distinguished parent-field that
references another scope object. (The global scope object's parent-field
is \lstinline|null|.) This linked list of scope objects is called a
\emph{scope chain}.  The value of an identifier \lstinline|x| is the
value of the first \lstinline|x|-field in the \emph{current scope
chain}.  When a new variable \lstinline|y| is defined, the field
\lstinline|y| is added to the scope object at the head of the scope
chain.

Since scope objects are ordinary JavaScript objects, JavaScript's
\lstinline|with| statement lets us add arbitrary
objects to the scope chain.  Given the features discussed below, which include
\lstinline|with|,
it is not clear whether JavaScript is lexically scoped.  In
this section, we describe how JavaScript's scope-manipulation statements
are desugared into $\lambda_{JS}$, which is obviously lexically scoped.

\subsubsection{Local Variables}

In JavaScript, functions close over their current scope chain
(intuitively, their static environment).  Applying a closure sets the
current scope chain to be that in the closure.  In addition, an empty
scope object is added to the head of the scope chain.  The function's
arguments and local variables (introduced using \lstinline|var|) are
properties of this scope object.

Local variables are automatically \emph{lifted} to the top of the function.
As a result, in a fragment such as this---
\begin{lstlisting}
function foo() {
  if (true) { var x = 10 }
  return x }

foo() $\twoheadrightarrow$ 10
\end{lstlisting}
---the \lstinline|return| statement has access to the variable that
appears to be defined inside a branch of the \lstinline|if|.  This
can result in somewhat unintuitive answers:
\begin{lstlisting}
function bar(x) {
  return function() {
    var x = x;
    return x }}

bar(200)() $\twoheadrightarrow$ undefined
\end{lstlisting}
Above, the programmer might expect the \lstinline|x| on the right-hand side of
\lstinline|var x = x| to reference the argument \lstinline|x|. However,
due to lifting, all bound occurrences of \lstinline|x| in the nested
function reference the local variable \lstinline|x|. Hence,
\lstinline|var x = x| reads and writes back the initial value of \lstinline|x|.
The initial value of local variables is
\lstinline|undefined|.

We can easily give a lexical account of this behavior.  A local
variable declaration, \lstinline|var x = e|, is desugared to an
\lstset{language=LambdaJS}
assignment, \lstinline|x = e|.  Furthermore, we add a let-binding
at the top of the enclosing function:
\begin{lstlisting}
let (x = ref undefined) $\cdots$
\end{lstlisting}

\subsubsection{Global Variables}

\lstset{language=JavaScript}
Global variables are subtle.  Global variables are properties of
the global scope object (\lstinline|window|), which has
a field that references itself:
\begin{lstlisting}
window.window === window $\twoheadrightarrow$ true
\end{lstlisting}
Therefore, a program can obtain a reference to the global scope object
by simply referencing \lstinline|window|.\footnote{In addition,
  \lstinline|this| is bound to \lstinline|window| in function
  applications (\reffig{f:classes}).}

As a consequence, globals seem to break lexical scope, since we can
observe that they are properties of \lstinline|window|:
\begin{lstlisting}
var x = 0;
window.x = 50;
x $\twoheadrightarrow$ 50
x = 100;
window.x $\twoheadrightarrow$ 100
\end{lstlisting}
However, \lstinline|window| is the only scope object that is directly
accessible to JavaScript programs~\cite[Section 10.1.6]{ecma262}.  We
maintain lexical scope by abandoning global variables.  That is, we
simply desugar the obtuse code above to the following:
\begin{lstlisting}
window.x = 0;
window.x = 50;
window.x $\twoheadrightarrow$ 50
window.x = 100;
window.x $\twoheadrightarrow$ 100
\end{lstlisting}
Although global variables observably manipulate \lstinline|window|,
local variables are still lexically scoped.  We can thus reason about local
variables using substitution, $\alpha$-renaming, and other standard
techniques.

\subsubsection{With Statements} The \lstinline|with| statement is a
widely-acknowledged JavaScript wart.  A \lstinline|with| statement
adds an arbitrary
object to the front of the scope chain:
\begin{lstlisting}
function(x, obj) {
  with(obj) {
    x = 50; // if obj.x exists, then obj.x = 50, else x = 50
    return y } } // similarly, return either obj.y, or window.y
\end{lstlisting}
We can desugar \lstinline|with| by turning the comments above into code:
\begin{lstlisting}
function(x, obj) {
  if (obj.hasOwnProperty("x")) { obj.x = 50 }
  else { x = 50 }
  if ("y" in obj) { return obj.y }
  else { return window.y } }
\end{lstlisting}
Nested \lstinline|with|s require a little more care, but can be dealt with
in the same manner. However, desugaring \lstinline|with| is
non-compositional.  We will return to this point in
\refsect{s:scaling}.

\paragraph{What are Scope Objects?}

Various authors (including ourselves) have developed JavaScript tools
that work with a subset of JavaScript that is intuitively lexically scoped
(e.g.,~\cite{anderson:inference,adsafe,fbjs,guha:intrusiondetection,heidegger:recency,miller:caja}).
We show how JavaScript can be desugared into lexically scoped $\lambda_{JS}$,
validating these assumptions.
As a result, we no longer need scope objects in the specification;
they may instead be viewed as an implementation
strategy.\footnote{Scope objects are especially well suited for implementing
  \lstinline|with|.  Our desugaring strategy for \lstinline|with| increases
  code-size linearly in the number of nested \lstinline|with|s, which 
  scope-objects avoid.}

\subsection{Type Conversions and Primitive Operators}
\label{s:prims}

\begin{figure}[t]
\begin{displaymath}
\begin{array}{rcl}
e & =    & \cdots 
    \mid   \js{$op_n$($e_1 \cdots e_n$)} \\
E & =   & \cdots
    \mid  op_n\js{(}v \cdots E\ e \cdots\js{)} \\
E' & =     & \cdots
     \mid  op_n\js{(}v \cdots E' e \cdots\js{)}
\\
\delta_n & : & op_n \times v_1 \cdots v_n \rightarrow c + err
\end{array}
\end{displaymath}

\infax[E-Prim]
{\js{$op_n$($v_1 \cdots v_n$)} \hookrightarrow \delta_n(op_n, v_1 \cdots v_n)}

\medskip\hrule
\caption{Primitive Operators}
\label{f:prims}
\end{figure}

JavaScript is not a pure object language.  We can observe the
difference between primitive numbers and number objects:
\begin{lstlisting}
x = 10;
y = new Number(7)
typeof x $\twoheadrightarrow$ "number"
typeof y $\twoheadrightarrow$ "object"
\end{lstlisting}
Moreover, JavaScript's operators include implicit type
conversions between primitives and corresponding objects:
\begin{lstlisting}
x + y $\twoheadrightarrow$ 17
\end{lstlisting}
We can redefine these type conversions without changing objects' 
values:
\begin{lstlisting}
Number.prototype.valueOf = function() { return 0 }
x + y $\twoheadrightarrow$ 10
y.toString() $\twoheadrightarrow$ "7"
\end{lstlisting}
Both
\lstinline|+| and \lstinline|*| perform implicit coercions, and \lstinline|+|
also concatenates strings:
\begin{lstlisting}
x + y.toString() $\twoheadrightarrow$ "107" // 10 converted to the string "10"
x * y.toString() $\twoheadrightarrow$ 70  // "7" converted to the number 7
\end{lstlisting}

This suggests that JavaScript's operators are complicated.  Indeed, the
standard specifies \lstinline|x + y| with a 15-step 
algorithm~\cite[Section 11.6.1]{ecma262}
that refers to three pages of metafunctions.  Buried in these details
are four primitive operators: primitive addition, string concatenation,
and number-to-string and string-to-number type coercions.

These four primitives are essential and intuitive.  We therefore model
them with a conventional $\delta$ function (\reffig{f:prims}).  The
remaining details of operators are type-tests and method
invocations; as the examples above suggest, JavaScript internally
performs operations such as \lstinline|y.valueOf()| and
\lstinline|typeof x|.  In $\lambda_{JS}$ we
make these type-tests and method calls explicit.

This paper does not enumerate all the primitives that $\lambda_{JS}$
needs.  Instead, the type of $\delta$ constrains their behavior
significantly, which often lets us reason without a specific $\delta$
function. (For instance, due to the type of $\delta$, we know that primitives
 cannot manipulate the heap.)

\section{Soundness and Adequacy of $\lambda_{JS}$}
\label{s:testing}
\lstset{language=LambdaJS}

\paragraph{Soundness}
We mechanize $\lambda_{JS}$ with {\sc plt} Redex~\cite{felleisen:redex}.  The
 process of
mechanizing helped us find errors in our semantics, particularly in
the interactions of control operators (\reffig{f:control}).  We use our
mechanized semantics to test~\cite{klein:redextesting} $\lambda_{JS}$ for
safety.
\begin{progress}[Progress]
If $\sigma e$ is a closed, well-formed configuration, then either:
\begin{itemize}

\item $e \in v$,

\item $e =\ $\lstinline|err $v$|, for some $v$, or

\item $\sigma e \rightarrow \sigma' e'$, where $\sigma' e'$ is a closed,
  well-formed configuration.
\end{itemize}
\end{progress}
This property requires additional evaluation rules for runtime type
errors, and definitions of well-formedness.  We elide them from the
paper, as they are conventional.  The supplemental material contains
these details.

\begin{figure}[t]
\makebox[\textwidth]{
\xymatrix{
\textrm{JavaScript Program} 
  \ar[rr]^{\textrm{\emph{desugar}}} 
  \ar[dd]_{\textrm{Real Implementations}} & &
\textrm{$\lambda_{JS}$ Program}
  \ar[dd]^{\textrm{$\lambda_{JS}$ interpreter}} \\
& & \\
\texttt{stdout} 
  \ar@{<->}[rr]_{\texttt{diff}} & &
\texttt{stdout}
}
}
\medskip\hrule
\caption{Testing Strategy for $\lambda_{JS}$}
\label{f:testing}
\end{figure}
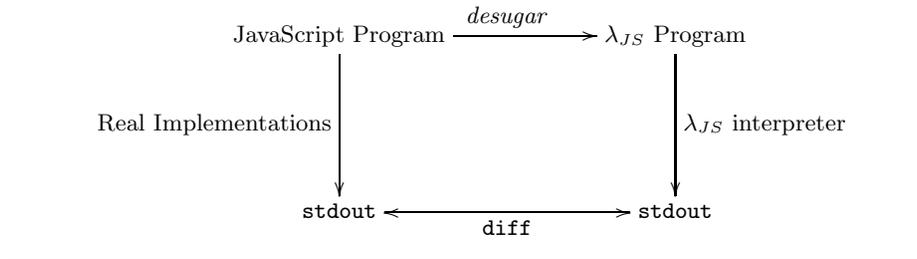

\begin{figure}[t]
\lstset{language=JavaScript}
\center
\begin{tabular}{|l|c|}
\hline
Syntactic Form & Occurrences (approx.) \\
\hline
\lstinline|with| blocks & \hphantom{0}15 \\
\lstinline|var| statements & 500 \\
\lstinline|try| blocks & \hphantom{0}20 \\
\lstinline|function|s & 200 \\
\lstinline|if| and \lstinline|switch| statements & 90 \\
\lstinline|typeof| and \lstinline|instanceof| & \hphantom{0}35 \\
\lstinline|new| expressions & \hphantom{0}50 \\
\lstinline|Math| library functions & \hphantom{0}15 \\
\hline
\end{tabular}

\medskip\hrule
\caption{Test Suite Coverage}
\label{f:coverage}
\end{figure}

\paragraph{Adequacy} $\lambda_{JS}$ is a semantics for the core of
JavaScript.  We have described how it models many aspects
of the language's semantics, warts and all.  Ultimately, however, a
small core language has limited value to those who want to reason
about programs written in full JavaScript.

Given our method of handling JavaScript via desugaring, we are
obliged to show that desugaring
and the semantics enjoy two properties.  First, we must show that
all JavaScript programs can be desugared to $\lambda_{JS}$.
\begin{desugarclaim1}[Desugaring is Total]
For all JavaScript programs $e$, $desugar\llbracket e \rrbracket$ is defined.
\end{desugarclaim1}
Second, we must demonstrate that our semantics corresponds to what
JavaScript
implementations actually do.
\begin{desugarclaim2}[Desugar Commutes with Eval]
  For all JavaScript programs $e$, $desugar\llbracket
  eval_{JavaScript}(e) \rrbracket =
  eval_{\lambda_{JS}}(desugar\llbracket e\rrbracket)$.
\end{desugarclaim2}
\lstset{language=JavaScript}
We could try to prove these claims, but that just begs the question:
What is $eval_{JavaScript}$?  A direct semantics
would require evidence of its own adequacy.

In practice, JavaScript is truly defined by its major implementations.
Open-source Web browsers are accompanied by extensive JavaScript test
suites.  These test suites help the tacit standardization of JavaScript
across major implementations.\footnote{For example, the Firefox
  JavaScript test suite is also found in the Safari source.}  We use
these test suites to \emph{test} our semantics.

\Reffig{f:testing} outlines our testing strategy.  We first define an
interpreter for $\lambda_{JS}$.  This is a straightforward exercise;
the interpreter is a mere 100 
LOC, and easy to inspect since it is
based directly on the
 semantics.\footnote{{\sc plt} Redex can evaluate expressions in 
a mechanized semantics. However, our tests are too large for Redex's
evaluator.}
Then, for any JavaScript program, we
should be able to run it both directly and in our semantics.  For direct
execution we employ three JavaScript implementations: SpiderMonkey 
(used by Firefox), V8 (used by Chrome), and Rhino (an implementation in Java).
  We desugar the same program into
$\lambda_{JS}$ and run the result through our interpreter.  We then
check whether  our $\lambda_{JS}$ interpreter produces the
same output as each JavaScript implementation.

Our tests cases are a significant portion of the Mozilla JavaScript
test suite.  We omit the following tests:
\begin{itemize}
\item  Those that target Firefox-specific JavaScript extensions.
\item  Those that use \lstinline|eval|.
\item Those that target library details, such as regular expressions.
\end{itemize}
The remaining tests are about 5,000 LOC unmodified.

Our $\lambda_{JS}$ interpreter produces exactly the same output as
Rhino, V8, and SpiderMonkey on the entire test suite.
  \Reffig{f:coverage} indicates that these tests 
employ many interesting syntactic forms, including statements like
\lstinline|with| and \lstinline|switch| that are considered
complicated.  We make the following observations:
\begin{itemize}
\item No prior semantics for JavaScript accounts for all these forms
  (e.g., Maffeis et al.~\cite{maffeis:jssemantics} do not model
  \lstinline|switch|).

\item We account for much of JavaScript by desugaring.  Therefore, these
  tests validate both our core semantics and our desugaring strategy.

\item These tests give us confidence that our implemented tools are
  correct.
\end{itemize}

\section{Example: Language-Based Sandboxing}
\label{s:isolation}

\lstset{language=JavaScript}
Web platforms often combine programs from several different sources on
the same page.  For instance, on a portal page like iGoogle, a user
can combine a weather widget with a stock ticker widget; on Facebook,
users can run applications.  Unfortunately, this means programs
from different authors can in principle examine data from one another,
which creates the possibility that a malicious application may
steal data or create other harm.  To prevent both accidents and
malice, sites must somehow sandbox widgets.

To this end, platform developers have defined safe sub-languages (often called
``safe subsets'') of JavaScript like ADsafe~\cite{adsafe}, 
Caja~\cite{miller:caja}, and
 Facebook JavaScript
(\FBJS)~\cite{fbjs}.  These are designed as sub-languages---rather than as
whole new languages with, perhaps, security types---to target
developers who already know how to write JavaScript Web applications.
These sub-languages disallow blatantly dangerous features such as
\lstinline|eval|.  However, they also try to establish
more subtle security properties using syntactic restrictions, as well as
runtime checks that they insert into untrusted code.  Naturally, this
raises the question whether these sub-languages function as
advertised.

Let us consider the following property, which is inspired by \FBJS\
and Caja: we wish to prevent code in the sandbox from communicating
with a server.  For instance, we intend to block the
\XMLHttpRequest\ object:
\begin{lstlisting}
var x = new window.XMLHttpRequest()
x.open("GET", "/get_confidential", false)
x.send("");
var result = x.responseText
\end{lstlisting}
For simplicity, we construct a sub-language that only disallows access
to
\XMLHttpRequest.  A complete solution would use our techniques to block
other communication mechanisms, such as \lstinline|document.write| and
\lstinline|Element.innerHTML|.

We begin with short, type-based proofs that exploit the
compactness of $\lambda_{JS}$.  We then use our tools to migrate
from $\lambda_{JS}$ to JavaScript.

\subsection{Isolating JavaScript}

\lstset{language=JavaScript}

We must precisely state ``disallow access to
\XMLHttpRequest''.  In JavaScript, \lstinline|window.XMLHttpRequest|
references the \XMLHttpRequest\ constructor, where \lstinline|window|
names the global object.  We make two assumptions:
\begin{itemize}
\item In $\lambda_{JS}$, we allocate the global object at location
  \lstinline|0|.  This is a convenient convention that is easily ensured
  by desugaring.

\item The \XMLHttpRequest\ constructor is only accessible
  as a property of the global object.  This assumption is valid as long
  as we do not use untrusted libraries (or can analyze their code).
\end{itemize}
Given these two assumptions, we can formally state ``disallow access to
\XMLHttpRequest'' as a property of $\lambda_{JS}$ programs:
\lstset{language=LambdaJS}
\begin{safety1}[Safety]\label{def:safety}
  $e$ is safe if $e \not = E\langle\langle$
\lstinline|deref (ref 0)|$\rangle$
\lstinline|[$\qq$XMLHttpRequest$\qq$]|$\rangle$.
\end{safety1}
Note that in the definition above, the active expression is
\lstinline|(deref (ref 0))|, and the evaluation context is $E\langle
\hole\js{[$\qq$XMLHttpRequest$\qq$]}\rangle$.

Intuitively, ensuring safety appears to be easy.  Given an untrusted
$\lambda_{JS}$ program, we can elaborate property accesses,
\lstinline|$e_1$[$e_2$]|, to
\lstinline|$lookup$($e_1$,$e_2$)|, where $lookup$ is defined
in \Reffig{f:lookup}.

\begin{figure}[t]
\begin{lstlisting}
$lookup$ = func(obj, field) {
  return if (field === $\qq$XMLHttpRequest$\qq$) { undefined }
         else { (deref obj)[field] }
}
\end{lstlisting}
\medskip\hrule
\caption{Safe Wrapper for $\lambda_{JS}$}
\label{f:lookup}
\end{figure}

This technique\footnote{Maffeis
  et al.'s blacklisting~\cite{maffeis:isolating}, based on
  techniques used in {\FBJS}, has this form.}
has two problems.  First, 
this elaboration does not allow access to the
\lstinline|$\qq$XMLHttpRequest$\qq$| property of \emph{any} object.
Second,
although $lookup$ may appear ``obviously correct'',  the actual
wrapping in Caja, {\FBJS}, and other sub-languages occurs in JavaScript,
not in a core calculus like $\lambda_{JS}$. Hence, $lookup$
does not directly correspond to any JavaScript function.  We could
write a JavaScript function that resembles $lookup$, but it
would be wrought with various implicit type conversions and method
calls (\refsect{s:prims}) that could break its intended behavior.
Thus, we start with safety for $\lambda_{JS}$ before tackling
JavaScript's details.

\subsection{Types for Securing $\lambda_{JS}$}
\label{s:xhrsafety}

\begin{figure}[t]
$T = \JS$

\infax[T-String]{ \Gamma \vdash string : \JS }

\infrule[T-Id]
{\Gamma(x) = T}
{\Gamma \vdash x : T}

\infrule[T-Fun]
{\Gamma, x_1 : \JS, \cdots, x_n : \JS \vdash e : \JS}
{\Gamma \vdash \js{func ($x_1\cdots x_n$) \{ return e \}} : \JS}

\infrule[T-Prim]
{\Gamma \vdash e_1 : \JS \andalso \cdots \andalso
 \Gamma \vdash e_n : \JS \andalso}
{\Gamma \vdash \delta_n(op_n, e_1\cdots e_n) : \JS}

The type judgments for remaining forms are similar to \textsc{T-Prim} and
\textsc{T-Fun}: namely, $\Gamma \vdash e : \JS$ if all subexpressions of
$e$ have type $\JS$.  However, \lstinline|$e_1$[$e_2$]| is \emph{not typable}.

\medskip\hrule
\caption{Type System that Disallows Field Lookup}
\label{f:xhr-silly}
\end{figure}

\begin{figure}[t]

$T = \cdots \mid \Safe$

\infax[Sub-Safe]{\Safe <: \JS}

\infrule[T-Sub]
{\Gamma \vdash e : S \andalso S <: T}
{\Gamma \vdash e : T}

\infrule[T-SafeValue]
{ v \ne \js{$\qq$XMLHttpRequest$\qq$}}
{ \Gamma \vdash v : \Safe}

\infrule[T-GetField]
{\Gamma \vdash e_1 : \JS \andalso \Gamma \vdash e_2 : \Safe}
{\Gamma \vdash \js{$e_1$[$e_2$]} : \JS}

\infrule[T-IfSafe]
{x \in dom(\Gamma) \andalso
 \Gamma \vdash e_2 : \JS \andalso
 \Gamma[x : \Safe] \vdash e_3 : \JS}
{\Gamma \vdash
  \js{if ($x\ $ === $\qq$XMLHttpRequest$\qq$) \{ $\ e_2\ $\} else \{$\ e_3\ $ \}} : \JS}

\medskip\hrule
\caption{Type System for Blocking Access to \XMLHttpRequest}
\label{f:xhr-types}
\end{figure}

\begin{figure}[t]

\infrule[T-IfTrue-XHR]
{\Gamma \vdash e_2 : \JS}
{\Gamma \vdash 
\js{if ($\qq$XMLHttpRequest$\qq$ === $\qq$XMLHttpRequest$\qq$) \{ $\ e_2\ $ \} else \{ $\ e_3\ $\}} : \JS}

\infrule[T-IfTrue]
{\Gamma \vdash e_2 : \JS}
{\Gamma \vdash 
 \js{if (true) \{ $\ e_2\ $ \} else \{ $\ e_3\ $ \}} : \JS}

\medskip\hrule
\caption{Auxiliary Typing Rules for Blocking Access to \XMLHttpRequest}
\label{f:xhr-types-aux}
\end{figure}

Our goal is to determine whether a $\lambda_{JS}$ program is safe
(\refdef{def:safety}).  We wish to do this without making unnecessary
assumptions.  In particular, we do not assume that $lookup$
(\reffig{f:lookup}) is itself safe.

We begin by statically disallowing \emph{all} field accesses.  The trivial type
system in \Reffig{f:xhr-silly} achieves this, since it excludes a typing
rule for \lstinline|$e_1$[$e_2$]|.  This type system does not catch
conventional type errors.  Instead, it has a single type,
$\JS$, of statically safe JavaScript expressions
(\refdef{def:safety}).  The following theorem is evidently true:
\begin{safety1thm}\label{thm:safety1}
  For all $\lambda_{JS}$ expressions $e$, if $\cdot \vdash e : T$ and $e
  \twoheadrightarrow e'$ then $e'$ is safe.
\end{safety1thm}

We need to extend our type system to account for $lookup$, taking care
not to violate \refthm{thm:safety1}.  Note that $lookup$ is currently
untypable, since field access is untypable.  However, the conditional in
$lookup$ seems to ensure safety; our goal is to prove that it does.
\
Our revised type system is shown in \reffig{f:xhr-types}.  The new type,
$\Safe$, is for expressions that provably do not evaluate to the
string \lstinline|$\qq$XMLHttpRequest$\qq$|. Since primitives like
string concatenation yield values of type $\JS$ (\textsc{T-Prim} in 
\reffig{f:xhr-silly}), programs cannot manufacture unsafe strings 
with type $\Safe$.  (Of course, trusted primitives could yield values
of type $\Safe$.)

Note this important peculiarity: \emph{These new typing rules are
purpose-built for $lookup$}.  There are other ways to establish safe
access to fields.  However, since we will rewrite all expressions
\lstinline|$e_1$[$e_2$]| 
to \lstinline|$lookup$($e_1$,$e_2$)|, our type system need
only account for the syntactic structure of $lookup$.

Our revised type system admits $lookup$, but we must prove
\refthm{thm:safety1}.  It is sufficient to prove the following
lemmas:\footnote{Additional proof details are in the supplemental material.}
\begin{safety1lem1}[Safety]\label{lem:safety}
  If $\cdot \vdash e : \JS$, then $e \ne
  E\langle\js{v[$\qq$XMLHttpRequest$\qq$]}\rangle$, for any value $v$.
\end{safety1lem1}
The proof of this lemma is by induction on typing derivations, given the
typing rules in \reffig{f:xhr-silly} and \reffig{f:xhr-types}.  This lemma
also holds for the typing rules in \reffig{f:xhr-types-aux}, which we
introduce below.
\begin{safety1lem2}[Subject Reduction]\label{lem:subset-preservation}
  If $\cdot \vdash e : \JS$, and $e \rightarrow e'$, then $\cdot \vdash
  e' : \JS$.
\end{safety1lem2}

\paragraph{Proof Technique} The typing rules for $lookup$ (\reffig{f:xhr-types})
require a technique introduced in \emph{occurrence typing} for Typed
Scheme~\cite{thf:typedscheme}.

Although $lookup$ is typable, subject reduction requires all
expressions in this reduction sequence to be typable:
\begin{lstlisting}
$lookup$($window$, $\qq$XMLHttpRequest$\qq$)
$\rightarrow$ if ($\qq$XMLHttpRequest$\qq$ === $\qq$XMLHttpRequest$\qq$) { undefined }
   else { (deref $window$)[$\qq$XMLHttpRequest$\qq$] }
$\rightarrow$ if (true) { undefined }
   else { (deref $window$)[$\qq$XMLHttpRequest$\qq$] }
$\rightarrow$ undefined
\end{lstlisting}
The intermediate expressions above are not typable, although they are
intuitively safe.  We can make them typable by extending our type
system with the typing rules in \reffig{f:xhr-types-aux}, which let us
prove subject reduction.

However, we have to ensure that our new typing rules do not violate
safety (\reflem{lem:safety}).  Intuitively, \reflem{lem:safety} still
holds, since our newly-typable expressions are not of the form
\lstinline|$v$[$\qq$XMLHttpRequest$\qq$]|.

Our type system may appear ad hoc, but it simply reflects the nature of
JavaScript security solutions.  Note that our type system is merely a
means to an end: the main result is the conclusion of
\refthm{thm:safety1}, which is a property of the runtime semantics.

\subsection{Scaling to JavaScript}
\label{s:scaling}

\lstset{language=JavaScript}
Since we can easily implement a checker for our type
system,\footnote{The supplemental material includes a 150-line
implementation.} we might claim we have a result for JavaScript as
follows: desugar JavaScript into $\lambda_{JS}$ and type-check the
resultant $\lambda_{JS}$ code.  This strategy is, however,
unsatisfying because seemingly harmless changes to a typable JavaScript program
may result in a program that fails to type-check, due
to the effects of desugaring.  This would make the language appear
whimsical to the widget developer.

Instead, our goal is to define a safe sub-language (just as, say, Caja
and {\FBJS} do).  This safe sub-language would provide syntactic
safety criteria, such as:
\begin{itemize}

\item The JavaScript expression \lstinline|$e_1$ + $e_2$| is safe when
  its subexpressions are safe.

\item \lstinline|$e_1$[$e_2$]|, when rewritten to
  \lstinline|$lookup$($e_1$, $e_2$)|, is safe, but fails if $e_2$
  evaluates to \lstinline|"XMLHttpRequest"|.
\end{itemize}

Our plan is as follows.  We focus on the \emph{structure} of the
desugaring rules and show that a particular kind of compositionality
in these rules suffices for showing safety.  We illustrate this
process by extending the $\lambda_{JS}$ result to include JavaScript's addition
(which, as we explained in
\refsect{s:prims}, is non-trivial).  We then generalize this
process to the rest of the language.

\subsubsection{Safety for Addition}

By \refthm{thm:safety1}, it is sufficient to determine whether $\Gamma \vdash
desugar \llbracket \js{$e_1$ + $e_2$} \rrbracket : \JS$.  Proving
this, however, would benefit from
some constraints on $e_1$ and $e_2$.  Consider the following
proposition:
\begin{prop1}\label{prop:prop1}
  If $\Gamma \vdash desugar \llbracket e_1 \rrbracket : \JS$ and $\Gamma
  \vdash desugar \llbracket e_2 \rrbracket : \JS$, then $\Gamma \vdash
  desugar \llbracket$\lstinline|$e_1$ + $e_2$|$\rrbracket : \JS$.
\end{prop1}
By \reflem{lem:safety}, this proposition entails that if $e_1$ and
$e_2$ are safe, then $e_1 + e_2$ is safe.  But is the proposition
true?  $desugar\llbracket \js{$e_1$ + $e_2$} \rrbracket$ produces an
unwieldy $\lambda_{JS}$ expression with explicit type-conversions and
method calls.  Still, a quick inspection of our implementation shows
that:
\lstset{language=LambdaJS}
\begin{lstlisting}
$desugar\llbracket e_1\texttt{ + }e_2\rrbracket$ = let (x = $desugar\llbracket e_1\rrbracket$) let (y = $desugar\llbracket e_2\rrbracket$) $\cdots$
\end{lstlisting}
$desugar\llbracket e_1\texttt{ + }e_2\rrbracket$ simply recurs on its
subexpressions and does not examine the result of $desugar\llbracket
e_1\rrbracket$ and $desugar\llbracket e_2\rrbracket$.  Moreover, the
elided body does not contain additional occurrences of $desugar\llbracket
e_1\rrbracket$ and $desugar\llbracket e_2\rrbracket$.  Thus, we can write
the right-hand side as a two-holed \emph{program context}:\footnote{Due
  to lack of space, we
  do not formally define program contexts for $\lambda_{JS}$ in this
  paper, but evaluation contexts offer a strong hint.}
\begin{lstlisting}
$desugar\llbracket e_1\texttt{ + }e_2\rrbracket$ = $C_+\langle desugar\llbracket e_1\rrbracket, desugar\llbracket e_2\rrbracket \rangle$
$C_+$ = let (x = $\hole_1$) let (y = $\hole_2$) $\cdots$
\end{lstlisting}
Therefore, desugaring \lstinline|$e_1$ + $e_2$| is \emph{compositional}. 

A simple replacement lemma~\cite{wright:types} holds for our type system:
\begin{replacement}[Replacement]\label{lem:replacement}
If:
\begin{itemize}

\item[i.] $\mathcal{D}$ is a deduction concluding $\Gamma \vdash C\lbrack
  e_1, e_2\rbrack : \JS$,

\item[ii.] Subdeductions $\mathcal{D}_1, \mathcal{D}_2$ prove that
  $\Gamma_1 \vdash e_1 : \JS$ and $\Gamma_2 \vdash e_2 : \JS$ respectively,

\item[iii.] $\mathcal{D}_1$ occurs in $\mathcal{D}$, at the position
  corresponding to $\hole_1$, and $\mathcal{D}_2$ at the
  position corresponding to $\hole_2$, and

\item[iv.] $\Gamma_1 \vdash e_1' : \JS$ and $\Gamma_2 \vdash e_2' : \JS$,

\end{itemize}
then $\Gamma \vdash C\langle e_1', e_2'\rangle : \JS$.
\end{replacement}

Replacement, along with weakening of environments,
gives us our final lemma:

\begin{plusreplacement}\label{lem:plusreplacement}
If:
\begin{itemize}

\item $x : \JS, y : \JS \vdash C_+\lbrack x, y \rbrack : \JS$, and

\item $\Gamma \vdash desugar\llbracket e_1\rrbracket : \JS$ and $\Gamma
  \vdash desugar\llbracket e_2 \rrbracket : \JS$,

\end{itemize}
then $\Gamma \vdash C_+ \langle desugar\llbracket e_1\rrbracket,
desugar\llbracket e_2 \rrbracket \rangle : \JS$.
\end{plusreplacement}
The conclusion of \reflem{lem:plusreplacement} is the conclusion of
\refprop{prop:prop1}.  The second hypothesis of
\reflem{lem:plusreplacement} is the only hypothesis of
\refprop{prop:prop1}.  Therefore, to prove \refprop{prop:prop1}, we
simply need to prove $x : \JS, y : \JS \vdash C_+\langle x, y \rangle :
\JS$.

\lstset{language=JavaScript}
We establish this using our tools.  We assume \lstinline|x| and
\lstinline|y| are safe (i.e., have type \JS), and desugar and
type-check the expression
\lstinline|x + y|.  Because this succeeds, the machinery above---in
particular, the replacement lemma---tells us that we may admit
\lstinline|+| into our safe sub-language.

\subsubsection{A Safe Sub-Language}

The proofs of lemma~\ref{lem:replacement} and
\ref{lem:plusreplacement} do not rely on the definition of $C_+$.
For each construct, we must thus ensure that the desugaring rule can
be written as a program context, which we easily verify by inspection.
We find this true for all syntactic forms other than
\lstinline|with|, which we omit from our safe sub-language (as do
other sub-language such as Caja and {\FBJS}).
  If \lstinline|with| were considered
important, we could extend our machinery to determine what
circumstances, or with what wrapping, it too could be considered safe.

Having checked the structure of the desugaring rules, we must still
establish that their expansion does no harm.  We mechanically populate
a type environment with placeholder variables, create expressions of
each kind, and type-check. All forms pass type-checking, except for the 
following:
\begin{itemize}
\item 
  \lstinline|x[y]| and \lstinline|x.XMLHttpRequest| do not
  type---happily, as they are unsafe!  This is
  acceptable because these unsafe forms will be wrapped in $lookup$.

\item However, \lstinline|x[y]++|, \lstinline|x[y]--|,  \lstinline|++x[y]|,
  and  \lstinline|--x[y]|
  also fail to type due to the structure of code they generate on
  desugaring. Yet, we believe these forms are safe; we could account
  for them with additional typing rules, as employed below for $lookup$.
\end{itemize}

\subsubsection{Safety for $lookup$}

As \refsect{s:xhrsafety} explained, we designed our type system to
account for $lookup$ (\reffig{f:lookup}).  However, $lookup$ is in
$\lambda_{JS}$, whereas we need a  wrapper in
JavaScript.  A direct translation of $lookup$ into JavaScript yields:
\begin{lstlisting}
$lookupJS$ = function(obj, field) {
 if (field === "XMLHttpRequest") { return undefined }
 else { return obj[field] } }
\end{lstlisting}
Since $lookupJS$ is a closed expression that is inserted as-is into
untrusted scripts, we can desugar and type-check it in isolation.
Doing so, however, reveals a surprise: $desugar
\llbracket lookupJS \rrbracket$ does not type-check.

When we examine the generated $\lambda_{JS}$ code, we see that
\lstinline|obj[field]| is desugared into an expression that explicitly
converts \lstinline|field| to a string. (Recall that field names are
strings.)  If, however, \lstinline|field| is itself an object, this
conversion includes the method call \lstinline|field.toString()|.
Working backward, we see that the following exploit would succeed:
\begin{lstlisting}
$lookupJS$(window, { toString: function() { return "XMLHttpRequest" } })
\end{lstlisting}
where the second argument to $lookupJS$ (i.e., the expression in the
field position) is a literal object that has a single method,
\lstinline|toString|, which returns 
\lstinline|"XMLHttpRequest"|.
Thus, not only does $lookupJS$ not type, it truly is unsafe!

Our type system successfully caught a bug in our JavaScript
implementation of $lookup$.  The fix is simple: ensure
that \lstinline|field| is a primitive string:
\begin{lstlisting}
$safeLookup$ = function(obj, field) {
  if (field === "XMLHttpRequest") { return undefined }
  else if (typeof field === "string") { return obj[field] }
  else { return undefined } }
\end{lstlisting}
This code truly is safe, though to prove it we need to extend our type
system.  We design the extension by studying the result of
desugaring $safeLookup$.\footnote{Desugaring produces 200 LOC of 
pretty-printed $\lambda_{JS}$.  We omit this code from the paper, but it is
available online.}

We have noted that desugaring  evinces the unsafe method call.
 However, \lstinline|toString| is called only if
\lstinline|field| is not a primitive. This conditional is inserted
\emph{by desugaring}:
\lstset{language=LambdaJS}
\begin{lstlisting}
if (typeof field === $\qq$location$\qq$) { ... field.toString() ... }
else { field }
\end{lstlisting}\pagebreak
Thus, the second \lstinline|if| in $safeLookup$ desugars to:
\begin{lstlisting}
if (typeof field === $\qq$string$\qq$) { 
  obj[if (typeof field === $\qq$location$\qq$) { ... field.toString() ... }
      else { field }] }
\end{lstlisting}
To now reach \lstinline|field.toString()|, both conditions must hold.
Since this cannot happen, the unsafe code block is unreachable.

Recall, however, that we designed our type system for $\lambda_{JS}$
around the syntactic structure of the lookup guard.  With this more
complex guard, we must extend our type system to employ
if-splitting---which we already used in \refsect{s:xhrsafety}---a
second time.  As long as our extension does not violate safety
(\reflem{lem:safety}) and subject reduction
(\reflem{lem:subset-preservation}), the arguments in this section
still hold.

\subsection{Perspective}

\lstset{language=JavaScript}
In the preceding sections, we rigorously developed a safe sub-language of
JavaScript that disallows access to \XMLHttpRequest.  In addition, we
outlined a proof of correctness for the runtime ``wrapper''.
To enhance
isolation, we have to disallow access to a few other properties, such
as \lstinline|document.write| and \lstinline|Element.innerHTML|.
Straightforward variants of the statements and proofs in this section could
verify such systems.  We believe our approach can scale to tackle more
sophisticated security properties as well.

Nevertheless, our primary goal in this section is not to define a
safe sub-language of JavaScript,
but rather to showcase our semantics and
tools:
\begin{itemize}

\item $\lambda_{JS}$ is small.  It is much smaller than other
  definitions and semantics for JavaScript.  Therefore, our proofs are
  tractable.

\item $\lambda_{JS}$ is adequate and tested.  This gives us confidence
  that our arguments are applicable to real-world JavaScript.

\item $\lambda_{JS}$ is conventional, so we are free to use standard
  type-soundness techniques~\cite{wright:types}.  In contrast, working with
  JavaScript's scope objects would be onerous.
  This section is littered with statements of the form
  $\Gamma \vdash e : \JS$.  Heap-allocated scope objects would
  preclude the straightforward use of $\Gamma$, thus complicating the
  proof effort (and perhaps requiring new techniques).

\item Finally, $desugar$ is compositional.  Although we developed a type
  system for $\lambda_{JS}$, we were able to apply our results to
  most of JavaScript by exploiting the compositionality of $desugar$.
\end{itemize}

\section{Related Work}
\label{s:related}

\paragraph{JavaScript Semantics} JavaScript is specified in 200 pages of
prose and pseudocode~\cite{ecma262}. This specification is barely
amenable to informal study, let alone proofs.  Maffeis, Mitchell, and
Taly~\cite{maffeis:jssemantics} present a 30-page operational semantics,
based directly on the JavaScript specification.  Their semantics covers
most of JavaScript directly, but does omit a few syntactic forms.

Our approach is drastically different. $\lambda_{JS}$ is a semantics for
the core of JavaScript, though we desugar the rest of JavaScript into
$\lambda_{JS}$.  In \refsect{s:testing}, we present evidence that our
strategy is correct.  $\lambda_{JS}$ and desugaring together are much
smaller and simpler than the semantics presented by Maffeis, et al.
Yet, we cover all of JavaScript (other than \lstinline|eval|) and
account for a substantial portion of the standard libraries as well
 (available in the supplementary material).

Maffeis, et al.~demonstrate adequacy by following the standard, though
they discuss various differences between the standard and
implementations.  In \refsect{s:testing}, we demonstrate adequacy by
running 3rd-party JavaScript tests in $\lambda_{JS}$ and comparing
results with mainstream JavaScript implementations.

A technical advantage of our semantics is that it is conventional.  For
example, we use substitution instead of scope objects (\refsect{s:scope}).
Therefore, we can use conventional techniques, such as subject
reduction, to reason in $\lambda_{JS}$.  It is unclear how to build
type systems for a semantics that uses scope objects.

David Herman~\cite{herman:classicjs} defines a {\sc ceks} machine for a small 
portion of
JavaScript.  This machine is also based on the
standard and inherits some of its complexities, such as implicit type
conversions.

CoreScript~\cite{yu:corescript} models an imperative subset of
JavaScript, along with portions of the {\sc dom}, but omits
essentials such as functions and objects.  Moreover, their big-step
semantics is not easily amenable to typical type safety proofs.

\paragraph{Object Calculi} $\lambda_{JS}$ is an untyped, object-based
language with prototype inheritance.  However, $\lambda_{JS}$ does not
have methods as defined in object calculi.  Without methods,
most object calculi cease to be interesting.  However, we do desugar
JavaScript's method invocation syntax to self-application in
$\lambda_{JS}$~\cite[Chapter 18]{abadi:objects}.

$\lambda_{JS}$ and JavaScript do not support cloning, which is a crucial
element of other prototype-based languages, such as
Self~\cite{ungar:self}.  JavaScript does support Self's prototype
inheritance, but the surface syntax of JavaScript does not permit direct
access to an object's prototype (\refsect{s:proto-stx}).  Without
cloning, and without direct access to the prototype, JavaScript
programmers cannot use techniques such as dynamic inheritance and
mode-switching~\cite{abadi:objects}.

\paragraph{Types for JavaScript}
There are various proposed type systems for JavaScript that are
accompanied by semantics. However, these semantics are only defined for
small subsets of JavaScript, not the language in its entirety.
For example, Anderson et al.~\cite{anderson:inference} develop a type
system and a type inference algorithm for $JS_0$, a subset 
that excludes prototypes and first-class functions.  Heidegger and
Thiemann's recency types~\cite{heidegger:recency} admit prototypes and
first-class functions, but omit assignment.
In contrast, we account for all of JavaScript 
(excluding \lstinline|eval|).

\section*{Acknowledgments}

We thank Matthias Felleisen, Mike Samuel, Peter Thiemann,
and the anonymous reviewers
for their careful comments on an earlier draft. We thank 
Sergio Maffeis, Leo Meyerovich, and John Mitchell
 for enlightening discussions. This work was
partially supported by the NSF.

\section*{Revision Log}

\begin{itemize}

  \item \textbf{October 3, 2015}. Jean-Baptiste Jeannin identified some type-setting
  errors.

  \item \textbf{[April 15, 2011}. The reduction rules \textsf{E-Break-Break} and 
  \textsf{Err-Break-Reduction} rules were missing. Without \textsf{E-Break-Break},
   the following term gets stuck:
   \begin{lstlisting}
   (label x (break y (break x 1)))
   \end{lstlisting}
   \textsf{E-Break-Break} is now in Figure 8 of the paper and 
   \textsf{Err-Break-Reduction} is in the supplemental code.

   Thanks to David Van Horn for catching this bug.

  \item \textbf{July 8, 2010}. Shriram Krishnamurthi and Rodolfo Toledo identified some
  type-setting errors.

  \item \textbf{April 24, 2010}. Jan Vitek identified some type-setting errors.

\end{itemize}

\bibliographystyle{abbrv}
\bibliography{thesis}

\begin{thebibliography}{10}

\bibitem{abadi:objects}
M.~Abadi and L.~Cardelli.
\newblock {\em A Theory of Objects}.
\newblock Springer-Verlag, 1996.

\bibitem{anderson:inference}
C.~Anderson, P.~Giannini, and S.~Drossopoulou.
\newblock Towards type inference for {JavaScript}.
\newblock In {\em European Conference on Object-Oriented Programming}, 2005.

\bibitem{borning:prototypes}
A.~Borning.
\newblock Classes versus prototypes in object-oriented languages.
\newblock In {\em ACM Fall Joint Computer Conference}, 1986.

\bibitem{chugh:sif}
R.~Chugh, J.~A. Meister, R.~Jhala, and S.~Lerner.
\newblock Staged information flow for {JavaScript}.
\newblock In {\em ACM SIGPLAN Conference on Programming Language Design and
  Implementation}, 2009.

\bibitem{adsafe}
D.~Crockford.
\newblock {ADSafe}.
\newblock \url{www.adsafe.org}.

\bibitem{ecma262}
{ECMAScript} language specification, 1999.

\bibitem{fbjs}
{Facebook}.
\newblock {FBJS}.
\newblock \url{wiki.developers.facebook.com/index.php/FBJS}.

\bibitem{felleisen:redex}
M.~Felleisen, R.~B. Findler, and M.~Flatt.
\newblock {\em Semantics Engineering with {PLT} {R}edex}.
\newblock {MIT} Press, 2009.

\bibitem{livshits:gatekeeper}
S.~Guarnieri and B.~Livshits.
\newblock {GateKeeper}: Mostly static enforcement of security and reliability
  policies for {JavaScript} code.
\newblock In {\em {USENIX} Security Symposium}, 2009.

\bibitem{guha:intrusiondetection}
A.~Guha, S.~Krishnamurthi, and T.~Jim.
\newblock Static analysis for {Ajax} intrusion detection.
\newblock In {\em International World Wide Web Conference}, 2009.

\bibitem{heidegger:recency}
P.~Heidegger and P.~Thiemann.
\newblock Recency types for dynamically-typed, object-based languages: Strong
  updates for {JavaScript}.
\newblock In {\em ACM SIGPLAN International Workshop on Foundations of
  Object-Oriented Languages}, 2009.

\bibitem{herman:classicjs}
D.~Herman.
\newblock {ClassicJavaScript}.
\newblock \url{www.ccs.neu.edu/home/dherman/javascript/}.

\bibitem{jensen:tajs}
S.~H. Jensen, A.~M{\o}ller, and P.~Thiemann.
\newblock Type analysis for {JavaScript}.
\newblock In {\em International Static Analysis Symposium}, 2009.

\bibitem{klein:redextesting}
C.~Klein and R.~B. Finder.
\newblock Randomized testing in {PLT Redex}.
\newblock In {\em ACM SIGPLAN Workshop on Scheme and Functional Programming},
  2009.

\bibitem{maffeis:jssemantics}
S.~Maffeis, J.~C. Mitchell, and A.~Taly.
\newblock An operational semantics for {JavaScript}.
\newblock In {\em Asian Symposium on Programming Languages and Systems}, 2008.

\bibitem{maffeis:isolating}
S.~Maffeis, J.~C. Mitchell, and A.~Taly.
\newblock Isolating {JavaScript} with filters, rewriting, and wrappers.
\newblock In {\em European Symposium on Research in Computer Security}, 2009.

\bibitem{miller:caja}
M.~S. Miller, M.~Samuel, B.~Laurie, I.~Awad, and M.~Stay.
\newblock {Caja}: Safe active content in sanitized {JavaScript}.
\newblock Technical report, Google Inc., 2008.
\newblock
  \url{http://google-caja.googlecode.com/files/caja-spec-2008-06-07.pdf}.

\bibitem{thf:typedscheme}
S.~Tobin-{H}ochstadt and M.~Felleisen.
\newblock The design and implementation of {T}yped {S}cheme.
\newblock In {\em ACM SIGPLAN-SIGACT Symposium on Principles of Programming
  Languages}, 2008.

\bibitem{ungar:self}
D.~Ungar and R.~B. Smith.
\newblock {SELF}: The power of simplicity.
\newblock In {\em ACM SIGPLAN Conference on Object-Oriented Programming
  Systems, Languages \& Applications}, 1987.

\bibitem{wright:types}
A.~Wright and M.~Felleisen.
\newblock A syntactic approach to type soundness.
\newblock {\em {Information and Computation}}, 115(1), 1994.

\bibitem{yu:corescript}
D.~Yu, A.~Chander, N.~Islam, and I.~Serikov.
\newblock Javascript instrumentation for browser security.
\newblock In {\em ACM SIGPLAN-SIGACT Symposium on Principles of Programming
  Languages}, 2007.

\end{thebibliography}

\end{document}